\documentclass[a4paper]{article}

\usepackage{multirow}
\usepackage{cite}
\usepackage{epsfig}
\usepackage{amsmath}

\usepackage{color}

\def\bea{\begin{eqnarray}}
\def\eea{\end{eqnarray}}
\def\pbpnull{\langle\bar{\psi_0}\psi_0\rangle}
\def\pbp{\langle\bar{\psi}\psi\rangle}
\def\gsim{\mathrel{\rlap{\lower4pt\hbox{\hskip1pt$\sim$}}\raise1pt\hbox{$>$}}}

\begin{document}

\hfill ITP-Budapest 658, WUB/02-05

\vspace{2cm}

\begin{center}{\Large\bf QCD thermodynamics with continuum}

\vspace{0.5cm}

{\Large\bf extrapolated Wilson fermions I.}\\

\vspace{2cm}

Szabolcs Bors\'anyi$^{\,a}$, Stephan D\"urr$^{\,ab}$, Zolt\'an Fodor$^{\,abc}$, 

\vspace{0.5cm}

Christian Hoelbling$^{\,a}$, S\'andor D. Katz$^{\,c}$, Stefan Krieg$^{\,ab}$, D\'aniel N\'ogr\'adi$^{\,c}$

\vspace{0.5cm}

K\'alm\'an K. Szab\'o$^{\,a}$, B\'alint C. T\'oth$^{\,a}$, Norbert Trombit\'as$^{\,c}$

\vspace{1cm}

\end{center}

\hspace{0.5cm}$^a${\em University of Wuppertal, Department of Physics, Wuppertal D-42097, Germany}

\vspace{0.2cm}

\hspace{0.5cm}$^b${\em J\"ulich Supercomputing Center, Forschungszentrum J\"ulich, J\"ulich D-52425, Germany}

\vspace{0.2cm}

\hspace{0.5cm}$^c${\em E\"otv\"os University, Institute for Theoretical Physics, Budapest 1117, Hungary}

\vspace{2cm}

\begin{abstract}

QCD thermodynamics is considered using Wilson fermions in the fixed scale approach. The temperature dependence
of the renormalized chiral condensate, quark number susceptibility and Polyakov loop is measured at four lattice spacings
allowing for a controlled continuum limit. The light quark masses are fixed to heavier than physical values in this
first study. Finite volume effects are ensured to be negligible by using approriately large box sizes. The final continuum results
are compared with staggered fermion simulations performed in the fixed $N_t$ approach. The same continuum renormalization
conditions are used in both approaches and the final results agree perfectly. 

\end{abstract}

\newpage

\section{Introduction}
\label{introduction}

As the early universe evolved a transition occured at temperatures $T\approx 150 - 200$~MeV, which is related to the 
spontaneous breaking of chiral symmetry in QCD. The nature of the QCD transition \cite{Aoki:2006we} affects 
our understanding of the history of the universe; see e.g. \cite{Schwarz:2003du}.

Extensive experimental work is currently being done with heavy ion collisions to study the QCD transition, most 
recently at the Relativistic Heavy Ion Collider, RHIC and at the Large Hadron Collider, LHC. Both for the 
cosmological transition and for RHIC/LHC, the net baryon densities are quite small, thus the baryonic chemical 
potentials $\mu$ are much less than the typical hadron masses, $\mu$ is below 50~MeV at RHIC, even smaller at LHC 
and negligible in the early universe. Thus, a calculation at $\mu=0$ is directly applicable to the cosmological 
transition and most probably also determines the nature and absolute temperature of the transition at RHIC/LHC. 
Therefore, we carry out our analysis at $\mu=0$. Given the far-reaching implications, it is desirable to perform this calculations in a
framework that is conceptually clean. For a pedagogical review on the QCD transition at $\mu=0$ and $\mu >0$ see e.g. 
\cite{Fodor:2009ax} and for the first continuum result at $\mu>0$ -- namely, for the curvature on the chemical 
potential vs. temperature plane, see \cite{Endrodi:2011gv}.

When we analyze the absolute scale or any other question related to the $T>0$ QCD transition for the physically 
relevant case two ingredients are quite important.

First of all, one should use physical quark masses. The nature of the transition strongly depends on the quark mass. 
Lattice studies and effective models showed that in the three flavor theory for small or large quark masses the 
transition is a first order phase transition, whereas for intermediate quark masses it is an analytic crossover. 
Since the nature of the transition influences the absolute scale $T_c$ of the transition 
-- its value, mass dependence, uniqueness etc. -- the use of physical quark masses is essential for the 
determination of $T_c$, too. The absolute scale then goes into all observables. Whereas it is relatively easy to 
reach the physical value of the strange quark mass $m_s$ in present day lattice simulations, it is much more 
difficult to work with physical up and down quark masses $m_{ud}$, because they are 
much smaller: $m_s/m_{ud}\approx 28$. In calculations with $m_s/m_{ud}$ smaller than 28 the strange quark mass is 
usually tuned to its approximate physical value, whereas the average up and down quark masses are larger than the 
physical value.

Secondly, the nature and other characteristics of the $T>0$ QCD transition are known to suffer from discretization 
errors \cite{deForcrand:2007rq,Endrodi:2007gc}. Let us mention one example which underlines the importance of 
removing these discretization effects by performing a controlled continuum extrapolation. The three flavor theory 
with a large, $a\approx 0.3$~fm lattice spacing and standard staggered action predicts a critical pseudoscalar mass 
of about 300~MeV \cite{Karsch:2001nf}. This point separates the first order and cross-over regions. If we took another 
discretization, with another discretization error, the critical pseudoscalar mass turns out to be much smaller, well 
below the physical pion mass of 135~MeV. The only way to determine the physical features of the transition is to 
carry out a careful continuum limit analysis. It can be safely done only in the so-called scaling regime. This regime 
is reached when the lattice spacing $a$ is sufficiently small, smaller than some $a_{max}$. Dimensionless 
combinations of observables approach their continuum limit value (within their error bars) in the scaling regime with 
a correction term $c\cdot a^n$. Here $c$, $n$ and $a_{max}$ depend on the action and on the dimensionless combination. 
The values of $c$ and $a_{max}$ are typically 
unknown, whereas the form of the action and the observables provide the value for $n$, usually without performing any 
simulations. To carry out a controlled continuum extrapolation at least three lattice spacings in the scaling regime 
are needed. Two points will always lie on a two parameter $c\cdot a^n$ curve, independently whether the lattice 
spacings are smaller than $a_{max}$ or not; the third point indicates if one reached the scaling regime.

It is numerically very demanding to fulfill both conditions. There are only a few cases, for which this has been 
achieved. 
Within the staggered formalism there are full 
results such as the nature of the transition 
\cite{Aoki:2006we}, the transition temperature \cite{Aoki:2006br,Aoki:2009sc,Bazavov:2011nk}, equation of 
state \cite{Aoki:2009sc} and fluctuations \cite{Borsanyi:2011sw}.

At $\mu>0$ lattice computations are more costly, see e.g. \cite{Fodor:2001au}. However
the curvature of the phase line separating the hadronic and
quark-gluon phases is already determined in the continuum
limit \cite{Endrodi:2011gv}. Results for the possible critical point
only simulations on coarse lattices are available; see e.g. \cite{Fodor:2004nz}.

It is important to note that fulfilling the second condition without fulfilling the first one still leads to 
universal results. In other words continuum extrapolated results with non-physical quark masses are universal. 
Independently of the action, simulation algorithm, scale setting procedure, they provide the same answer once the 
quark mass is fixed which is a non-trivial issue, but can be done e.g. by fixing the pion to Omega and kaon to Omega 
mass ratios: $M_\pi/M_\Omega$ and $M_K/M_\Omega$. These results are not the same as they are for physical 
quark masses, but they are well defined and unique. Contrary to this universality, fulfilling the first condition 
(physical quark mass) but not the second one (continuum extrapolation) leads to non-universal, 
non-physical results. These results still have unknown discretization errors.

Once the available computational resources are not enough to fulfill both conditions it is more advisable to carry 
out calculations with non-physical quark masses but perform the continuum limit extrapolation. As we have seen such 
results are universal and can be cross-checked with other results obtained by other fermion formalisms, actions etc.
For some recent Wilson thermodynamics results see 
\cite{Umeda:2008bd, Ejiri:2009hq, Maezawa:2009nd, Bornyakov:2009qh, Umeda:2010ye, Maezawa:2011aa, Umeda:2012er}.

In this paper we determine the temperature dependencies of a couple of observables (chiral condensate, strange
susceptibility, Polyakov loop) in 2+1 flavor QCD. We use Wilson fermions with six steps of stout smearing and tree level
clover improvement in the quark sector and a tree level improved action in the gauge sector; for the details of the
action see \cite{Durr:2008rw}. Our pion is non-physical, its mass is about 545~MeV, see later for a detailed
discussion for the mass. 

The structure of the paper can be summarized as follows. After this brief introductory Section \ref{introduction} a discussion on the 
advantages and disadvantages of Wilson thermodynamics is presented in Section \ref{choiceofthe}. The main features of the action and 
run parameters are listed in Section \ref{simulationpoints}. Our choice of renormalization procedures for the various measured
quantities are summarized in Section \ref{renormalization}. The results are given in Section \ref{results}. In Section
\ref{summary} we summarize and provide an outlook.

\section{Choice of the fermion formalism}
\label{choiceofthe}

This paper presents the first of a series, which deals with lattice QCD thermodynamics using Wilson fermions. The final
goal is to describe several bulk observables as a function of the temperature all the way to the continuum
limit with physical quark masses or in other words by approaching pion masses of 135~MeV and kaon masses of 495~MeV. 
While the framework has the advantage of the sound conceptual status of the Wilson fermion formulation ({\it a}), and we are
already able to reach the continuum limit ({\it b}), the present work suffers from the
large computational cost ({\it c}) and is therefore confined to larger than physical pion masses ({\it d}).
Let us discuss these aspects in some detail.

{\it ad a.} The vast majority of the large scale lattice QCD thermodynamics projects has been carried out with 
staggered fermions. Staggered fermions are the cheapest formulation of lattice QCD and working with them turned out 
to be quite successful and provided many interesting results both at $T=0$ and $T>0$. In their original form they describe 
four degenerate fermions -- tastes -- and one has to take the square root or the fourth root of the fermion 
determinant to describe two fermions or one fermion, respectively. This procedure is somewhat unattractive and there 
has been an ongoing debate in the literature whether it leads to the proper universality class. Wilson fermions do 
not raise such theoretical questions. Furthermore, staggered fermions suffer from the so-called taste symmetry 
violation. It means that instead of the physical pseudo-Goldston bosons there is a tower of pseudoscalars whose 
masses are typically well beyond the physical pion or kaon masses. The physical spectrum is expected to be 
restored only in the continuum limit. In the last five years some of the authors of the present paper have carried out a large scale staggered 
thermodynamics program, nevertheless it is desirable to cross check those results by repeating the analyses with 
Wilson fermions. Its theoretical cleanness is the most important reason to perform a systematic study with Wilson 
fermions.

{\it ad b.} As we discussed earlier lattice results are unambiguous only in the continuum limit. To this end we carry 
out our analysis at four lattice spacings. As we will see the results scale quite nicely and the continuum behavior 
can be extracted. The results are in good agreement with the results obtained with another fermion discretization,
stout smeared staggered fermions. There is, however, one important conceptual difference between Wilson and 
staggered thermodynamics. In the staggered case the standard procedure is to take a given temporal extent $N_t$
to control the lattice spacing and use various gauge couplings, light and strange quark masses $\beta$, $m_{ud}$ 
and $m_s$ to change the temperature. The set of parameters $\beta$, $m_{ud}$ and $m_s$, for which the physical 
content is the same (e.g. the ratio of the kaon decay constant and the pseudoscalar masses: $f_K/M_\pi$ and 
$f_K/M_K$) are called lines of constant physics (LCP). The fixed scale approach of Wilson thermodynamics 
\cite{Umeda:2008bd} is in some sense the opposite. One takes a given $\beta$ to fix the lattice spacing and uses several 
$N_t$ values to scan the temperature. The reason for this choice is the difficulty with the additive mass 
renormalization within the Wilson formalism. This additive term makes it particularly difficult to give the LCPs with 
good numerical precision. The fixed scale approach ensures by definition that the physical content remains the same 
(same bare parameters) even if we change the temperature.

{\it ad c.} Wilson fermions are usually more expensive than staggered fermions. There are at least two reasons for 
that. First of all the basic calculational step, fermion matrix $\times$ spinor multiplication, has four times as many 
floating point operations for Wilson fermions as for staggered fermions. Secondly, the computational costs for the 
inversion of the fermion matrix strongly depends on its condition number. The condition number for the fermion matrix 
has a strict bound given by the quark mass for the staggered case, whereas for the same quark mass with Wilson 
fermions the condition number can be much larger. It depends not only on the mass but also on the gauge configuration 
and thus fluctuates more.

{\it ad d.} Today, the large computational costs allow to study only systems with larger than physical pion 
masses; larger mass means smaller condition number, thus smaller computational costs. As we discussed earlier it is 
more reasonable to carry out a study with larger than physical mass and extrapolate to the continuum limit than use 
smaller or even physical quark masses at one or two lattice spacings only. The reason for that is simple: continuum 
extrapolated results can be compared with those of other groups and/or lattice actions. A non-continuum result has 
still an unknown uncertainty due to cutoff effects. As we mentioned we compare 
our findings in this paper using Wilson fermions with our earlier results using staggered fermions. A nice agreement 
is found.

\section{Simulation points and techniques}
\label{simulationpoints}

In this section the details of the Wilson simulations are outlined. We also performed staggered simulations in order
to compare the continuum results of the two formulations. The techniques of the staggered simulations follow 
\cite{Aoki:2006br, Aoki:2009sc} and are shortly summarized too.

\subsection{Action}

The gauge action used for the calculations was the Symanzik tree level
improved gauge action \cite{Symanzik:1983dc, Luscher:1984xn}
\begin{equation}
S_G^\mathrm{Sym}  =  \beta\,\,\Big[\frac{c_0}{3}\,
\sum_\mathrm{plaq} {\rm Re\, Tr\,}(1-U_\mathrm{plaq})
\, + \frac{c_1}{3}\, \sum_\mathrm{rect} {\rm Re \, Tr\,}
(1- U_\mathrm{rect})\Big],
\end{equation}
with the parameters $c_0 = 5/3$ and $c_1= -1/12$. The action for the
femionic sector was the clover improved \cite{Sheikholeslami:1985ij}
Wilson action
\begin{equation}
S_F^{\mathrm{SW}} = S_F^\mathrm{W}-
\frac{c_{\mathrm{SW}}}{4}\,\sum_x\,\sum_{\mu,\nu}
\overline{\psi}_x\,\sigma_{\mu\nu}F_{\mu\nu,x}\,\psi_x\,,
\end{equation}
where $S_F^\mathrm{W}$ is the Wilson fermion action. Six steps of stout
smearing \cite{Morningstar:2003gk} with smearing parameter $\varrho=0.11$ were
used. 
The clover coefficient was set to
its tree level value, $c_{\mathrm{SW}}=1.0$, which, for this type of smeared
fermions, essentially leads to an ${\cal O}(a)$ improved action
\cite{Hoffmann:2007nm} with improved chiral properties \cite{Capitani:2006ni}.

\subsection{Simulation algorithm}

The bare masses of the $u$ and $d$ quarks were taken to be degenerate,
therefore the configurations were generated using an $N_f=2+1$ flavor
algorithm. The light quarks were implemented via the Hybrid Monte Carlo (HMC)
algorithm~\cite{Duane:1987de}, whereas the strange quark was implemented using
the Rational Hybrid Monte Carlo (RHMC) algorithm~\cite{Clark:2006fx}. In order
to speed up the molecular dynamics calculations, the Sexton-Weingarten
multiple time-scale integration scheme \cite{Sexton:1992nu} combined with the
Omelyan integrator \cite{Takaishi:2005tz} was employed.  When all four extents
of the lattice were even, the usage of even-odd preconditioning
\cite{DeGrand:1988vx} gave an additional speed up factor of 2. For further
details on the algorithm see \cite{Durr:2008rw}.

\subsection{Simulation points}

\begin{table}
\begin{center}
\begin{tabular}{|c|c|c|c|c|}
\hline
$\beta$ & $a m_{ud}$ & $a m_s$  & $N_s$ & $N_t$ \\
\hline
\hline
3.30 &  -0.0985  &   -0.0710 & 32 & 4 - 16, 32 \\
\hline
3.57 &  -0.0260  &   -0.0115 & 32 & 4 - 16, 64 \\
\hline
3.70 &  -0.0111  &   0.0 & 48 & 8 - 28, 48 \\
\hline
3.85 &  -0.00336 &   0.0050  & 64 & 12 - 28, 64 \\
\hline
\end{tabular}
\end{center}
\caption{Simulation parameters. The $N_t$ values used for the finite temperature runs and the values used
for the zero temperature runs are separated by a comma.}
\label{tab:parameters}
\end{table}

The calculations were performed at four different gauge couplings,
$\beta=3.30$, $3.57$, $3.70$ and $3.85$. Only dimensionless ratios are measured 
and every dimensionful quantity is made dimensionless by appropriate powers of $m_\Omega$ or $m_\pi$.
The lattice spacing is also set by the physical value $m_\Omega = 1672~MeV$ and the 
four gauge couplings correspond to $a = 0.139(1),\; 0.093(1),\; 0.070(1)$ and $0.057(1)$ fm, respectively. 
At all four gauge couplings the bare masses are tuned such that $m_\pi / m_\Omega = 0.326(4)$
and $m_K/m_\Omega = 0.366(4)$ are constant, which means that $m_\pi \approx 545$ MeV and $m_K \approx 614$ MeV.
We tried to tune the $m_s/m_{ud}$ ratio to the same
value $m_s/m_{ud}=1.5$ as used in the staggered reference runs. We achieved
this within $2\%$ accuracy: the ratio $(2m_K^2-m_\pi^2)/m_\pi^2$ which in 
leading order chiral perturbation theory equals $m_s / m_{ud}$ is $1.530(7)$ for all four gauge couplings.
All four values of $m_s$ are such that if $m_{ud}$ is lowered to the physical point $m_K$ also becomes the
physical kaon mass.
The bare quark masses, spatial and
temporal lattice extents are shown in table \ref{tab:parameters} while
the measured masses are shown in table \ref{tab:masses}.  In all four cases $m_\pi L_s \gsim 8$.
At each finite temperature point around 1000 equilibrated trajectories were generated while around 500 at zero temperature points.

\subsection{Staggered simulations}

The goal of the staggered simulations detailed below was to provide a basis of
comparison for the Wilson data. We used the staggered action with two steps
of stout smearings ($\varrho=0.15$) as in most our thermodynamics studies e.g.
\cite{Aoki:2006br}. The line of constant physics used there was extended
to the fine lattices
in \cite{Aoki:2009sc,Borsanyi:2010bp,Borsanyi:2010cj}. We used the
bare strange masses for each given beta as were determined there. The light
quark mass was simply set to $2/3$ of the strange mass to achieve the
desired pion/kaon mass ratio.  The lattice spacing in these staggered
thermodynamics papers was set through the kaon decay constant.
For each used gauge coupling we determined the scale directly with zero
temperature runs at the light/strange mass ratio $2/3$. Using the most
precise scale setting observable we had access to, the $w_0$ scale 
\cite{Borsanyi:2012zs}, we checked that our original scale function
was still correct. We proceeded to compute the mass ratios $m_\pi/m_\Omega\simeq 0.32$
and $M_K/m_\Omega\simeq 0.36$ that turned out to agree with the Wilson values
at the level of $3\%$ in both cases.

For the renormalization we used dedicated
zero temperature runs with parameters precisely matching those in the finite
temperature simulations. We determined the vacuum condensate
$\langle\bar\psi\psi\rangle_0$ and the static potential $V(r)$, which
we needed for obtaining the multiplicative renormalization of the Polyakov
loop: $Z=\exp(V(x)/2)$. The obtained finite renormalized Polyakov loop
is then further transformed to our scheme (see section \ref{subsectpol}) by a finite renormalization factor.
One can select any physical scale $x$ to remove all divergences. Since
$w_0$ is our most accurately known scale, we used $x=\sqrt{8}w_0$, which is
approximately $0.5~fm$. We found that in the continuum, at this mass
ratio we have $m_{\Omega}w_0=1.466(15)$.

Our finite temperature staggered simulations were performed on $18^3\times 6$,
$24^3\times 8$, $32^3\times 10$ and $36^3\times 12$ lattices. After
performing the necessary steps of renormalization a continuum
extrapolation was carried out. For this extrapolation we used a cubic spline interpolation
(using roughly every second temperatures as node points). Then for every
temperature we evaluated several possible extrapolations (linear in $a^2$).
The choices included using all four lattice spacings or just three of them,
and also making the extrapolation for the reciprocal observable. The width
of the weighted histogram of the continuum extrapolations define a systematic
error, which we added to the statistical error in quadrature. The final
continuum extrapolation includes an additional overall percent-level error in
the temperature axis.

\begin{table}
\begin{center}
\begin{tabular}{|c|c|c|c|c|c|c|c|}
\hline
$\beta$ & $m_\pi/m_\Omega$ & $m_K/m_\Omega$  &  $m_s/m_{ud}$ & $a m_{PCAC}$ & $a m_\Omega$    & $a\; [fm]$ & $Z_A$ \\
\hline                                                                                      
\hline                                                                                      
3.30 &   0.332(3)          &   0.373(3)      & 1.529(2)    & 0.0428(2)     & 1.16(1)        &  0.139(1)     & 0.892(7)\\
\hline                                                                                       
3.57 &   0.319(6)          &   0.359(4)      & 1.531(2)    & 0.02649(4)    & 0.777(9)       &  0.093(1)     & 0.951(2) \\
\hline                                                                                      
3.70 &   0.326(5)          &   0.369(5)      & 1.531(3)    & 0.01994(4)    & 0.586(8)       & 0.070(1)      & 0.966(2) \\
\hline                                                                                      
3.85 &   0.314(7)          &   0.358(6)      & 1.528(4)    & 0.01559(2)    & 0.480(8)       & 0.057(1)      & 0.976(5) \\
\hline
\end{tabular}
\end{center}
\caption{Spectroscopy and $Z_A$ renormalization constant results from zero temperature simulations. 
The $m_s/m_{ud}$ column refers to $(2m_K^2-m_\pi^2)/m_\pi^2$. The lattice spacings are set by $m_\Omega = 1672$ MeV.}
\label{tab:masses}
\end{table}

\section{Renormalization}
\label{renormalization}

The bare chiral condensate is divergent in the continuum limit and both additive (power-like) and multiplicative (logarithmic)
divergences need to be removed. The resulting finite chiral condensate in the continuum limit can be compared with results
obtained using other regularizations for instance the staggered formulation since finite continuum quantities should not depend on
the regulator.

The additive and multiplicative renormalizations are treated separately in the following two subsections.
We follow \cite{Giusti:1998wy} which is based on \cite{Bochicchio:1985xa}. The measurement of the finite renormalization constant $Z_A$
is outlined in subsection \ref{subsectza}. The Polyakov loop also needs to be renormalized and our
scheme is defined below in subsection \ref{subsectpol}.

\subsection{Additive renormalization of the condensate}

On dimensional grounds the bare chiral condensate contains additive divergences of the type
\bea
\label{barepbp}
\frac{\pbpnull}{N_f} = c_0 + c_1(m_0-m_c) + c_2(m_0-m_c)^2 + \ldots
\eea
where $m_0$ is the dimensionful bare mass, $m_c$ is its critical value and $c_0$ is cubically, $c_1$ is quadratically and $c_2$ is
linearly divergent. The coefficients $c_i$ do not depend on the temperature hence they cancel in the quantity
\bea
\label{diff1}
\frac{\Delta_{\bar{\psi}\psi}(T)}{N_f} = \frac{\pbpnull(T) - \pbpnull(T=0)}{N_f}
\eea
just like in the staggered case. The cancellation is of order $O(a^{-3})$. 

However the $O(a^{-3})$ term in (\ref{barepbp}) can be explicitly
removed and the following quantity is free of cubic divergences \cite{Giusti:1998wy},
\bea
\label{pp}
\frac{\pbpnull}{N_f} - c_0 = 2 m_{PCAC} Z_A \int d^4 x \langle P_0(x) P_0(0) \rangle + \; \cdots\;,
\eea
but of course quadratically and linearly divergent pieces are still present. Here $P_0(x)$ is the bare pseudo-scalar condensate,
$m_{PCAC}$ is the PCAC mass and $Z_A(g_0)$ is a finite renormalization constant \cite{Giusti:1998wy}. Our conventions for the
definition of $P_0(x)$ and $\pbpnull$ are the same as in \cite{Giusti:1998wy}. Hence for the subtracted condensate we have,
\bea
\label{diff2}
\frac{\Delta_{\bar{\psi}\psi}(T)}{N_f} = 2 m_{PCAC} Z_A \Delta_{PP}(T) + \; \cdots\;,
\eea
where the short hand notation
\bea
\label{diffpp}
\Delta_{PP}(T) = \int d^4 x \langle P_0(x) P_0(0) \rangle(T) - \int d^4 x \langle P_0(x) P_0(0) \rangle(T=0)
\eea
was introduced. In (\ref{diffpp}) the cancellation between the finite and zero temperature terms is only
$O(a^{-2})$ however all additive divergences are still removed.

\subsection{Multiplicative renormalization of the condensate}

Multiplicative logarithmic divergences are still present in both (\ref{diff1}) and (\ref{diff2}).
Multiplying both expressions by first $Z_P$ and then by a renormalized mass $m_R = m_{PCAC} Z_A / Z_P$ leads to a renormalization group invariant
quantity $m_R\pbp_R$. Hence using (\ref{diff1}), (\ref{diff2}) and (\ref{diffpp}) we arrive at the expressions,
\bea
\label{finalfinal}
m_R \pbp_R(T) &=& 2 N_f m_{PCAC}^2 Z_A^2 \Delta_{PP}(T) \\
\label{final}
m_R \pbp_R(T) &=& m_{PCAC} Z_A \Delta_{\bar{\psi}\psi}(T) + \; \cdots\;,
\eea
where the former may be taken as the definition of $\pbp_R(T)$ and the latter agrees with it in the continuum limit. Both
are of course also finite. The finite cut-off corrections for (\ref{finalfinal}) is $O(a^2)$ at tree level 
provided the action is $O(a)$-improved, because the tree level improvement factors
for $m_R$ and $\pbp_R$ cancel in the product. The full non-perturbative improvement of $m_R$ and $\pbp_R$ we expect to be close to
tree level improvement because our smeared action with tree level improvement coefficient $c_{\mathrm{SW}}=1.0$ is very close to being
non-perturbatively $O(a)$-improved \cite{Hoffmann:2007nm, Capitani:2006ni}.
The quantities $\Delta_{\bar{\psi}\psi}(T)$ and $\Delta_{PP}(T)$ 
on the right hand sides are easy to measure on the
lattice and knowing $m_{PCAC}$ and $Z_A$ from zero temperature simulations allows one to define the finite and renormalization
group invariant continuum quantity $m_R \pbp_R(T)$ in two different ways. In the continuum limit the two definitions should
agree within errors.

On the other hand, using the first expression in (\ref{final}) to solve for $m_{PCAC} Z_A$ and substituting it into the second
expression leads to,
\bea
\label{final2}
m_R \pbp_R(T) = \frac{ \Delta_{\bar{\psi} \psi}^2(T) }{ 2 N_f \Delta_{PP}(T) } + \; \cdots\;,
\eea
which can directly be measured from the bare quantities $\Delta_{\bar{\psi}\psi}(T)$ and $\Delta_{PP}(T)$ without any need for measuring
$m_{PCAC}$ or the $Z_A$ renormalization factor. In \cite{Borsanyi:2011kg} the above ratio was used to define the chiral
condensate, in the present study we will use (\ref{finalfinal}) because we have found that this definition of the renormalized condensate scales best
among the three possible choices (\ref{finalfinal}), (\ref{final}) and (\ref{final2}).

\subsection{Computation of $Z_A$}
\label{subsectza}

We compute the renormalization constant of the local axial vector current $Z_A$ along the lines presented in \cite{Durr:2010vn,Durr:2010aw}. We generated independent sets of $N_f=3$ ensembles at four quark masses in the range $m_s/3<m_q<m_s$ and approximately equal physical volume $V\simeq (2~\mathrm{fm})^4$ at every $\beta$.

In a first step, we compute the local vector current renormalization constant $Z_V$ from the ratio
\begin{equation}
\xi(t)=\frac{\langle P_0(N_t/2)V_0(t)P_0(0)\rangle}{\langle P_0(N_t/2)P_0(0)\rangle}
\end{equation}
where 
\begin{equation}
P_0(t)=\int\,d^3xP_0(t,\vec{x})\;,
\qquad
V_0(t)=\int\,d^3x\,(\bar\psi_1\gamma_0\psi_1)(t,\vec{x})\;,
\end{equation}
are the zero 3-momentum projected bare pseudo-scalar and vector densities.
With tree level improvement one has \cite{Bakeyev:2003ff,Bhattacharya:2005rb}
\begin{equation}
Z_V(\beta,m) (1 + am) = \left(\xi(t_1)-\xi(t_2)\right)^{-1} \quad\mathrm{for}\quad 0<t_1< N_t/2<t_2<N_t
\end{equation}
We obtain $Z_A$ by using the standard RI-MOM procedure \cite{Martinelli:1994ty} with the improvement technique of \cite{Maillart:2008pv} to determine the ratio
\begin{equation}
\label{eq:za}
\frac{Z_A(\beta,m)}{Z_V(\beta,m)}=\frac{\Gamma_V(p)}{\Gamma_A(p)}
\end{equation}
from the off-shell amputated Greens functions $\Gamma_\Gamma(p)$. The dependence of the ratio (\ref{eq:za}) on the external
quark momenta $p$ is very mild and enters into our estimate of the systematic error. We linearly extrapolate the resulting
$Z_A(\beta,m)$ to $m=0$ to obtain the renormalization constant $Z_A(\beta)$, see table \ref{tab:masses}. 

\begin{figure}
\begin{center}
\includegraphics[width=8.5cm,height=6cm]{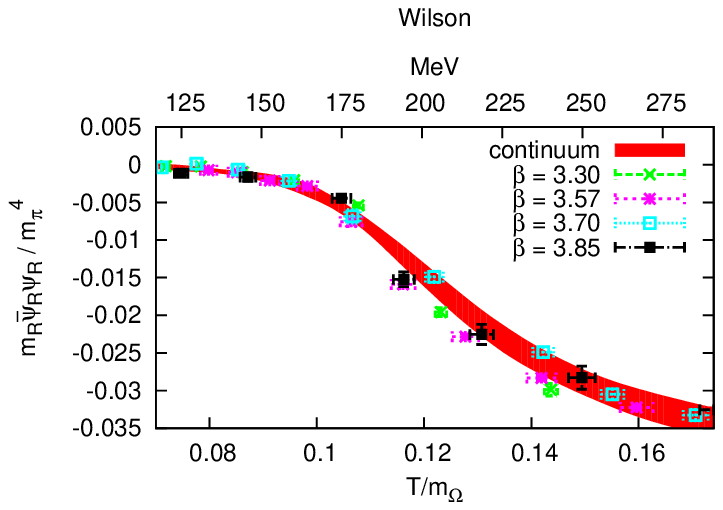} \hspace{-1.1cm} \includegraphics[width=7.5cm,height=6cm]{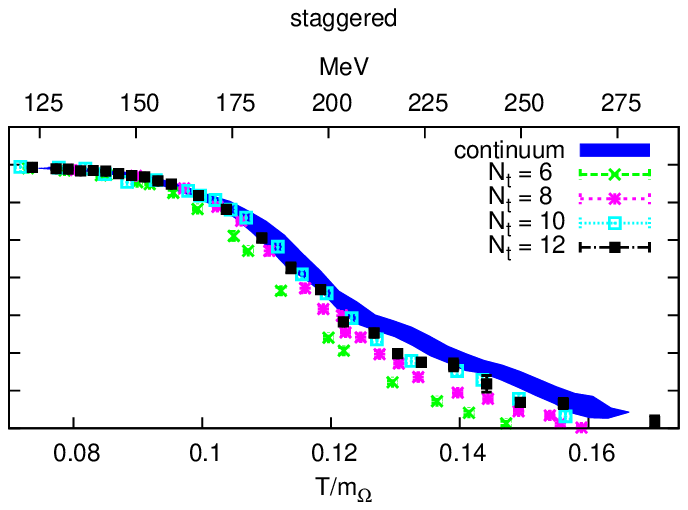} 

\vspace{1cm}

\includegraphics[width=12cm,height=8cm]{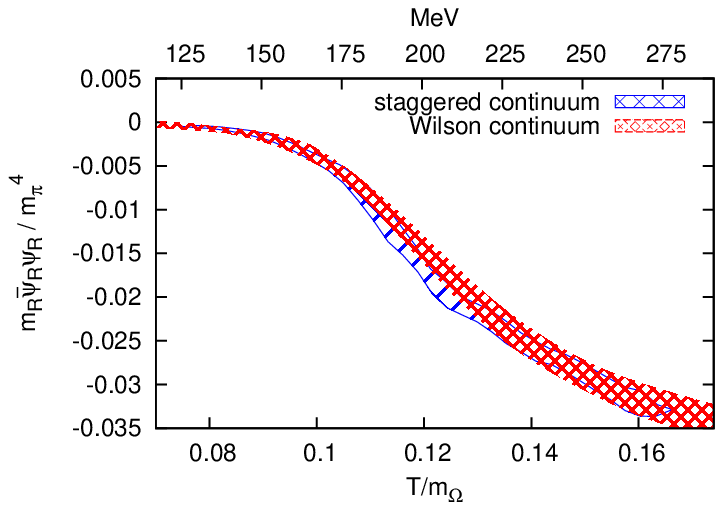}

\vspace{1cm}

\caption{Renormalized chiral condensate. The top left panel shows the Wilson results using the definition (\ref{finalfinal}) at
the four lattice spacings corresponding to four $\beta$ values. The continuum extrapolated result is also shown by the solid
band.  The top right panel shows the staggered results in the fixed-$N_t$ approach also together with the continuum extrapolated
result. The bottom panel compares the two continuum results.\label{figpbp}} \end{center} \end{figure}

\subsection{Polyakov loop}
\label{subsectpol}

The real part of the bare Polyakov loop also needs to be renormalized \cite{Aoki:2006br} in order to have a quantity with a finite continuum limit.
Since there is an additive divergence in the free energy, a convenient choice of renormalization prescription is demanding a fixed
value $L_*$ for the renormalized Polyakov loop at a fixed but arbitrary temperature $T_* > T_c$. Then the renormalized Polyakov loop $L_R$ is given
by
\bea
\label{pr2}
L_R(T) = \left( \frac{L_*}{L_0(T_*)} \right)^{\frac{T_*}{T}} L_0(T)
\eea
in terms of the bare Polyakov loop $L_0(T)$. We choose $T_* = 0.143 m_{\Omega}$ and $L_* = 1.2$. Other choices would simply
correspond to other renormalization schemes.

We imposed the same renormalization condition on the central value of our continuum extrapolated staggered data so they can be meaningfully compared.

\section{Results}
\label{results}

We have measured three quantities at each lattice spacing. The renormalized chiral condensate is 
sensitive to the remnant of the chiral transition whereas the renormalized Polyakov loop and the
strange quark number susceptibility are sensitive to the remnant of the confinement-deconfinement transition.

Each quantity is renormalized properly so that in the continuum limit finite and
regularization scheme independent values are obtained. The Wilson continuum
extrapolation is based on a cubic spline interpolation to temperatures not
reachable by the discrete range of $N_t$ and extrapolation $a \to 0$. We perform
global fits to all our data points including lattice spacing dependence in the fit
parameters as follows. Spline node points are randomly distributed along the
temperature axis. The parameters of our fit are the values of the observable at
the node points together with their lattice spacing dependence, i.e. $o_{1i}+c(a)o_{2i}$ 
for the $i^{th}$ node point. For the leading correction $c(a)$ we used two choices, $c(a)=a^2$ and
$c(a)=\alpha a$. We had 4-6 nodepoints with 1000 different random node point sets
each, resulting in $2\cdot 3\cdot 1000=6000$ fits alltogether. The results of the
fits are weighted with the fit qualities. For each temperature the median of these
fits is used as our final result and the systematic error comes from the central 68\%
of the distribution and the statistical error from a jackknife analysis. The
simulations were performed at 4 lattice spacings but not every observable is
measured in the full temperature range. For the temperature range where our
final results are presented we had at least 3 lattice spacings available in other words
at least 3 lattice spacings are used for the continuum extrapolation. The
obtained continuum results can then be compared with results from other
regularizations such as the staggered formulation.


Figure \ref{figpbp} shows the renormalized chiral condensate. The left panel contains the Wilson simulation results at four
lattice spacings together with the continuum extrapolated result. 
The right panel shows the chiral condensate result for the staggered
calculation at 4 fixed $N_t$ values 6, 8, 10 and 12 corresponding to four lattice spacings. Cubic spline is used to interpolate 
at fixed $N_t$ and these are then extrapolated to the continuum. The middle panel shows the two continuum extrapolated results. They are in
perfect agreement.

In an earlier exploratory work we used the definition (\ref{final2}) for the renormalized chiral condensate resulting in much larger
cut-off effects \cite{Borsanyi:2011kg}. The reason for using (\ref{finalfinal}) this time is exactly because the cut-off effects are
smaller as is visible from figure \ref{figpbp} compared to figure 1 in \cite{Borsanyi:2011kg}.

\begin{figure}
\begin{center}
\includegraphics[width=8.5cm,height=6cm]{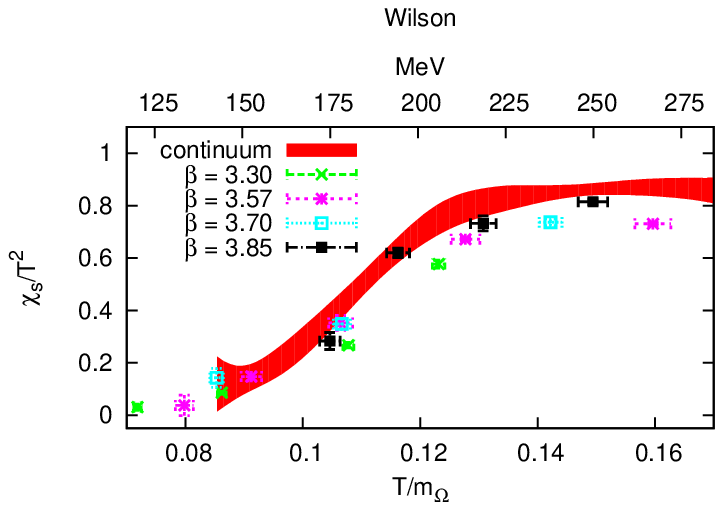} \hspace{-1.1cm} \includegraphics[width=7.5cm,height=6cm]{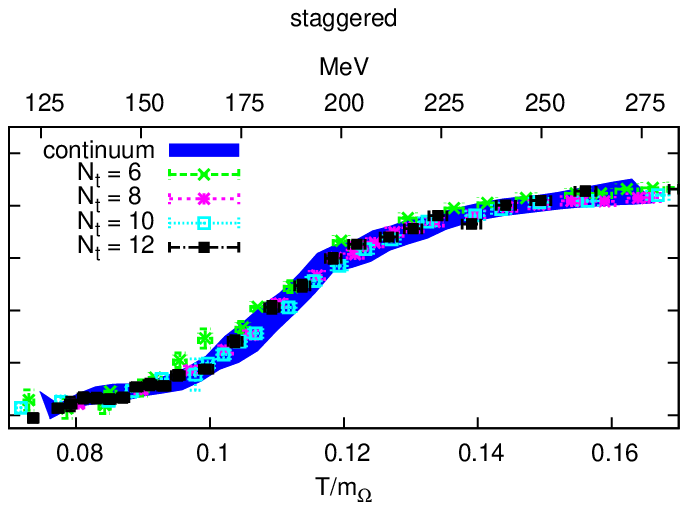} 

\vspace{2cm}

\includegraphics[width=12cm,height=8cm]{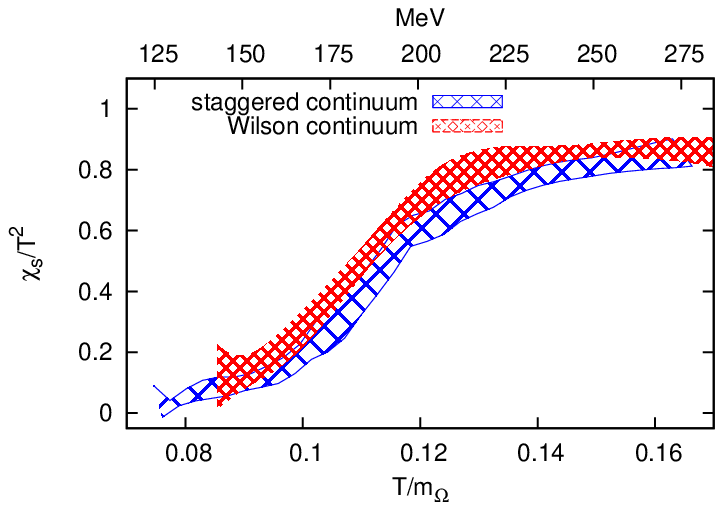}

\vspace{1cm}

\caption{Strange quark number susceptibility. The top left panel shows the Wilson results at the four lattice spacings corresponding to
four $\beta$ values. The continuum extrapolated result is also shown by the solid band. The top right panel shows the staggered results in
the fixed-$N_t$ approach also
together with the continuum extrapolated result. The bottom panel compares the two continuum results.\label{figqsusc}}
\end{center}
\end{figure}

The strange quark number susceptibility $\chi_s = T/V \, \partial^2 \log(Z) / \partial \mu_s^2$ is a 
sum of two contributions, the connected and disconnected terms. The disconnected part
is a very noisy quantity (as usual) and a large number of random vectors, $1200$, were needed in order to evaluate it precisely.
It is advantageous to generate the random vectors in the 12 diagonal spin-color blocks separately \cite{Ejiri:2009hq}.
Cut-off effects may be reduced by tree level improvement. The measured susceptibility is divided by its infinite volume, massless
Stefan-Boltzmann limit at the given finite $N_t$ temporal extent. The continuum limit is unchanged by this improvement since the
Stefan-Boltzmann limits are $1$ for $N_t \to \infty$; see table \ref{sb}.

\begin{table}
\begin{center}
\begin{tabular}{|c||c|c|c|c|c|c|c|c|c|c|}
\hline
$N_t$  &    8  &   10  &  12   &   14  &  16   &   18  &   20  &   22  &  24   &   26  \\
\hline
$SB$   & 1.522 & 1.265 & 1.161 & 1.110 & 1.081 & 1.062 & 1.049 & 1.040 & 1.034 & 1.028 \\
\hline
\end{tabular}
\end{center}
\caption{The tree level improvement factors of the strange quark number susceptibility for Wilson fermions. The values shown are the free, infinite
volume, massless Stefan-Boltzmann limits at given $N_t$. For $\beta = 3.30$ and $3.57$ only $N_t \geq 8$, 
for $\beta=3.70$ only $N_t \geq 12$ and for $\beta=3.85$ only $N_t\geq 14$ is used.}
\label{sb}
\end{table}

The results for the four lattice spacings are shown on figure \ref{figqsusc} together with the staggered results. 
What is plotted in both cases is the aforementioned tree level improved quark number susceptibility.

The number of data points is less for the quark number susceptibility than for the chiral condensate because for the condensate
odd $N_t$ values are used as well. In such a setup even-odd preconditioning can not be used resulting in a much slower simulation
and measurement. The large number of random vectors needed also forced us to also skip some even $N_t$ values in some cases only
having those that are divisible by 4.

Discretization errors are comparable to the chiral condensate. The results are shown on figure \ref{figqsusc} again with the
Wilson results on the left, staggered results on the right and the comparison of the two continuum results in the middle. Again
the agreement between the two continuum extrapolated results is perfect.

\begin{figure}
\begin{center}
\includegraphics[width=8.5cm,height=6cm]{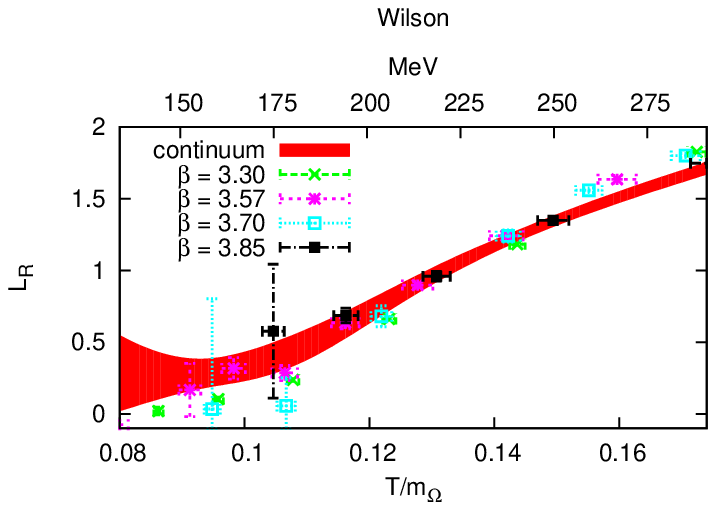} \hspace{-1.1cm} \includegraphics[width=7.5cm,height=6cm]{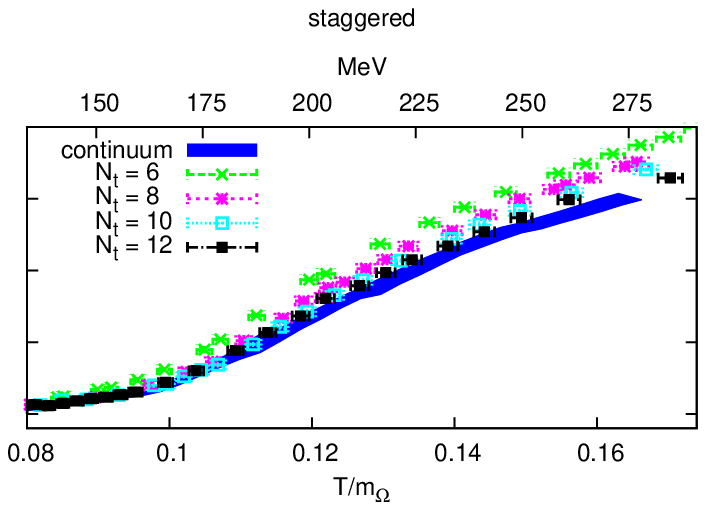} 

\vspace{2cm}

\includegraphics[width=12cm,height=8cm]{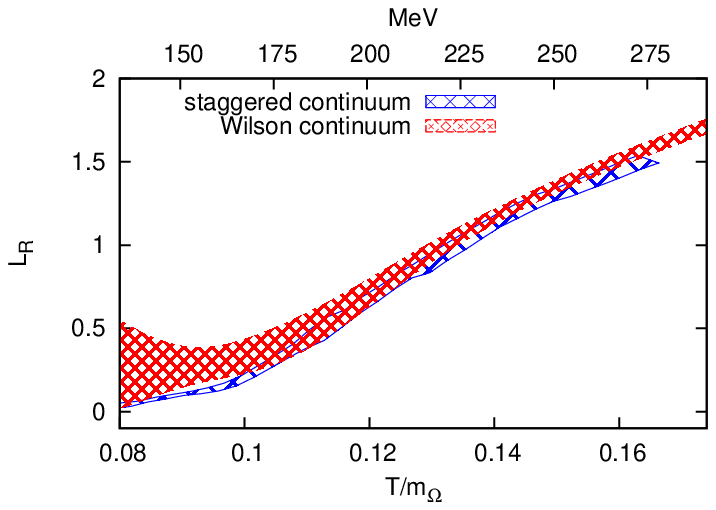}

\vspace{1cm}

\caption{Renormalized Polyakov loop. The top left panel shows the Wilson results at the four lattice spacings corresponding to
four $\beta$ values. The continuum extrapolated result is also shown by the solid band. The top right panel shows the staggered results in
the fixed-$N_t$ approach also together with the continuum extrapolated result. The bottom panel compares the two continuum results.\label{figploop}}
\end{center}
\end{figure}

The third quantity we measured is the renormalized Polyakov loop which is sensitive to the confinement-deconfinement transition
similarly to the quark number susceptibility. The results are shown on figure \ref{figploop} in the same format as before;
Wilson results on the left, staggered results on the right with the comparison of the two continuum extrapolated curves in the
middle. The bare Polyakov loop can be measured quite precisely but renormalization in our scheme at low temperatures causes the errors to
increase because of the division and high power in equation (\ref{pr2}). For higher temperatures the errors are smaller where the renormalization condition (\ref{pr2}) does not introduce large
uncertainties.

\section{Summary and outlook}
\label{summary}

In this work finite temperature QCD was studied using Wilson discretization for $2+1$ flavors of dynamical fermions. The unique
aspect of our work is the continuum extrapolation of all three renormalized quantities we have measured, chiral condensate,
strange quark number susceptibility and Polyakov loop, which is especially difficult with Wilson fermions. The difficulty lies in the
fact that the explicit chiral symmetry breaking of the Wilson formulation necessitates fine lattice spacings and consequently large
lattice volumes. The continuum extrapolation has been carried out for all temperatures and by using four lattice spacings.

The motivation for our work is twofold. First, it is desirable to obtain continuum QCD results from first principles which do not
contain theoretically not fully justified operations such as the fourth root trick of rooted staggered fermions. The Wilson
fermion formulation is theoretically sound and is known to be in the right universality class for QCD. Second, since a large body
of simulation results for both zero and finite temperatures exist with the staggered formulation
it is a worthwhile task to compare some of these continuum results with the continuum Wilson results. The
expectation is of course that if the staggered formulation is also in the right universality class the continuum results
will agree within errors.

The Wilson formulation is more expensive than the staggered formulation hence lowering the light quark masses towards their physical
value is much more difficult. In this work the pion mass has been set to $545\,MeV$, heavier than physical, but it is important to
note that at arbitrary quark masses the continuum results are universal. Hence provided the same renormalization prescription is
used for two discretizations the results should still agree in the continuum for arbitrary quark masses. We have carried out
staggered simulations using the same renormalization prescription as we did for the Wilson simulations and have compared the
continuum extrapolated results. Nice agreement was found for all three quantities.

In the future we plan to lower the light quark masses towards their physical value. With an unimproved fermion action this would
be almost hopeless however the use of stout improvement makes the fluctuation of low lying Dirac eigenvalues much less allowing
for a smaller pion mass at a given volume.

We also plan to compare the continuum extrapolated staggered and Wilson results with a formulation which combines the key
advantages of the two: chiral symmetry (staggered) and theoretical soundness (Wilson). Such a lattice chiral discretization is the
overlap formalism. Results at two lattice spacings are already available in the fixed-$N_t$ approach \cite{Borsanyi:2012xf}. Once completed
with a third or fourth lattice spacing a continuum extrapolation will be possible allowing for the comparison of three continuum
results from three discretizations of QCD.

\section*{Acknowledgment}

Computations were carried out on GPU clusters \cite{Egri:2006zm} at the
Universities of Wuppertal and Budapest as well as on supercomputers in
Forschungszentrum Juelich.

This work was supported by the EU Framework Programme 7 grant (FP7/2007-2013)/ERC No 208740 and by the Deutsche Forschungsgemeinschaft
grants FO 502/2 and SFB-TR 55.


\begin{thebibliography}{99}

\footnotesize

\bibitem{Aoki:2006we}
Y.~Aoki, G.~Endrodi, Z.~Fodor, S.~D.~Katz and K.~K.~Szabo,
Nature {\bf 443} (2006) 675
[arXiv:hep-lat/0611014].

\bibitem{Schwarz:2003du}
D.~J.~Schwarz,
Annalen Phys.\  {\bf 12}, 220 (2003)
[arXiv:astro-ph/0303574].

\bibitem{Fodor:2009ax}
Z.~Fodor and S.~D.~Katz,
arXiv:0908.3341 [hep-ph].

\bibitem{Endrodi:2011gv} 
G.~Endrodi, Z.~Fodor, S.~D.~Katz and K.~K.~Szabo,
JHEP {\bf 1104}, 001 (2011)
[arXiv:1102.1356 [hep-lat]].

\bibitem{deForcrand:2007rq}
P.~de Forcrand, S.~Kim and O.~Philipsen,
PoS {\bf LAT2007}, 178 (2007)
[arXiv:0711.0262 [hep-lat]].

\bibitem{Endrodi:2007gc}
G.~Endrodi, Z.~Fodor, S.~D.~Katz and K.~K.~Szabo,
PoS {\bf LAT2007}, 182 (2007)
[arXiv:0710.0998 [hep-lat]].

\bibitem{Karsch:2001nf}
F.~Karsch, E.~Laermann and C.~Schmidt,
Phys.\ Lett.\  B {\bf 520}, 41 (2001)
[arXiv:hep-lat/0107020].

\bibitem{Aoki:2006br}
Y.~Aoki, Z.~Fodor, S.~D.~Katz and K.~K.~Szabo,
Phys.\ Lett.\  B {\bf 643}, 46 (2006)
[arXiv:hep-lat/0609068].

\bibitem{Aoki:2009sc}
Y.~Aoki, S.~Borsanyi, S.~Durr, Z.~Fodor, S.~D.~Katz, S.~Krieg and K.~K.~Szabo,
JHEP {\bf 0906}, 088 (2009)
[arXiv:0903.4155 [hep-lat]].

\bibitem{Bazavov:2011nk} 
A.~Bazavov {\it et al.},
Phys.\ Rev.\ D {\bf 85}, 054503 (2012)
[arXiv:1111.1710 [hep-lat]].

\bibitem{Borsanyi:2011sw} 
S.~Borsanyi, Z.~Fodor, S.~D.~Katz, S.~Krieg, C.~Ratti and K.~Szabo,
JHEP {\bf 1201}, 138 (2012)
[arXiv:1112.4416 [hep-lat]].

\bibitem{Fodor:2001au} 
Z.~Fodor and S.~D.~Katz,
Phys.\ Lett.\ B {\bf 534}, 87 (2002)
[hep-lat/0104001].

\bibitem{Fodor:2004nz} 
Z.~Fodor and S.~D.~Katz,
JHEP {\bf 0404}, 050 (2004)
[hep-lat/0402006].

\bibitem{Umeda:2008bd}
T.~Umeda, S.~Ejiri, S.~Aoki, T.~Hatsuda, K.~Kanaya, Y.~Maezawa and H.~Ohno,
Phys.\ Rev.\  D {\bf 79}, 051501 (2009)
[arXiv:0809.2842 [hep-lat]].

\bibitem{Ejiri:2009hq}
S.~Ejiri {\it et al.}  [WHOT-QCD Collaboration],
Phys.\ Rev.\  D {\bf 82}, 014508 (2010)
[arXiv:0909.2121 [hep-lat]].

\bibitem{Maezawa:2009nd} 
Y.~Maezawa {\it et al.}  [WHOT-QCD Collaboration],
Nucl.\ Phys.\ A {\bf 830}, 247C (2009)
[arXiv:0907.4203 [hep-lat]].

\bibitem{Bornyakov:2009qh} 
V.~G.~Bornyakov, R.~Horsley, S.~M.~Morozov, Y.~Nakamura, M.~I.~Polikarpov, P.~E.~L.~Rakow, G.~Schierholz and T.~Suzuki,
Phys.\ Rev.\ D {\bf 82}, 014504 (2010)
[arXiv:0910.2392 [hep-lat]].

\bibitem{Umeda:2010ye} 
T.~Umeda {\it et al.}  [WHOT-QCD Collaboration],
PoS LATTICE {\bf 2010}, 218 (2010)
[arXiv:1011.2548 [hep-lat]].

\bibitem{Maezawa:2011aa} 
Y.~Maezawa, T.~Umeda, S.~Aoki, S.~Ejiri, T.~Hatsuda, K.~Kanaya and H.~Ohno,
arXiv:1112.2756 [hep-lat].

\bibitem{Umeda:2012er} 
T.~Umeda {\it et al.}  [WHOT-QCD Collaboration],
arXiv:1202.4719 [hep-lat].

\bibitem{Durr:2008rw}
S.~Durr {\it et al.},
Phys.\ Rev.\  D {\bf 79}, 014501 (2009)
[arXiv:0802.2706 [hep-lat]].

\bibitem{Symanzik:1983dc}
K.~Symanzik,
Nucl.\ Phys.\  B {\bf 226} (1983) 187.

\bibitem{Luscher:1984xn} 
M.~Luscher and P.~Weisz,
Commun.\ Math.\ Phys.\  {\bf 97}, 59 (1985)
[Erratum-ibid.\  {\bf 98}, 433 (1985)].

\bibitem{Sheikholeslami:1985ij}
B.~Sheikholeslami and R.~Wohlert,
Nucl.\ Phys.\  B {\bf 259} (1985) 572.

\bibitem{Morningstar:2003gk}
C.~Morningstar and M.~J.~Peardon,
Phys.\ Rev.\  D {\bf 69} (2004) 054501
[arXiv:hep-lat/0311018].

\bibitem{Hoffmann:2007nm}
R.~Hoffmann, A.~Hasenfratz and S.~Schaefer,
PoS {\bf LAT2007} (2007) 104
[arXiv:0710.0471 [hep-lat]].

\bibitem{Capitani:2006ni} 
S.~Capitani, S.~Durr and C.~Hoelbling,
JHEP {\bf 0611}, 028 (2006)
[hep-lat/0607006].

\bibitem{Duane:1987de}
S.~Duane, A.~D.~Kennedy, B.~J.~Pendleton and D.~Roweth,
Phys.\ Lett.\  B {\bf 195} (1987) 216.

\bibitem{Clark:2006fx}
M.~A.~Clark and A.~D.~Kennedy,
Phys.\ Rev.\ Lett.\  {\bf 98} (2007) 051601
[arXiv:hep-lat/0608015].

\bibitem{Sexton:1992nu}
J.~C.~Sexton and D.~H.~Weingarten,
Nucl.\ Phys.\  B {\bf 380} (1992) 665.

\bibitem{Takaishi:2005tz}
T.~Takaishi and P.~de Forcrand,
Phys.\ Rev.\  E {\bf 73} (2006) 036706
[arXiv:hep-lat/0505020].

\bibitem{DeGrand:1988vx} 
T.~A.~DeGrand,
Comput.\ Phys.\ Commun.\  {\bf 52}, 161 (1988).

\bibitem{Borsanyi:2010bp} 
S.~Borsanyi {\it et al.}  [Wuppertal-Budapest Collaboration],
JHEP {\bf 1009}, 073 (2010)
[arXiv:1005.3508 [hep-lat]].

\bibitem{Borsanyi:2010cj} 
S.~Borsanyi, G.~Endrodi, Z.~Fodor, A.~Jakovac, S.~D.~Katz, S.~Krieg, C.~Ratti and K.~K.~Szabo,
JHEP {\bf 1011}, 077 (2010)
[arXiv:1007.2580 [hep-lat]].

\bibitem{Borsanyi:2012zs} 
S.~Borsanyi {\it et al.},
arXiv:1203.4469 [hep-lat].

\bibitem{Giusti:1998wy}
L.~Giusti {\it et al.},
Nucl.\ Phys.\  {\bf B538 } (1999)  249-277.
[hep-lat/9807014].

\bibitem{Bochicchio:1985xa}
M.~Bochicchio {\it et al.},
Nucl.\ Phys.\  {\bf B262 } (1985)  331.

\bibitem{Borsanyi:2011kg} 
S.~Borsanyi, Z.~Fodor, C.~Hoelbling, S.~D.~Katz, S.~Krieg, D.~Nogradi, B.~C.~Toth and K.~K.~Szabo,
arXiv:1111.3500 [hep-lat].

\bibitem{Durr:2010vn} 
S.~Durr {\it et al.},
Phys.\ Lett.\ B {\bf 701}, 265 (2011)
[arXiv:1011.2403 [hep-lat]].

\bibitem{Durr:2010aw} 
S.~Durr {\it et al.},
JHEP {\bf 1108}, 148 (2011)
[arXiv:1011.2711 [hep-lat]].

\bibitem{Bakeyev:2003ff} 
T.~Bakeyev {\it et al.}  [QCDSF-UKQCD Collaboration],
Phys.\ Lett.\ B {\bf 580}, 197 (2004)
[hep-lat/0305014].

\bibitem{Bhattacharya:2005rb} 
T.~Bhattacharya, R.~Gupta, W.~Lee, S.~R.~Sharpe and J.~M.~S.~Wu,
Phys.\ Rev.\ D {\bf 73}, 034504 (2006)
[hep-lat/0511014].

\bibitem{Martinelli:1994ty} 
G.~Martinelli, C.~Pittori, C.~T.~Sachrajda, M.~Testa and A.~Vladikas,
Nucl.\ Phys.\ B {\bf 445}, 81 (1995)
[hep-lat/9411010].

\bibitem{Maillart:2008pv} 
V.~Maillart and F.~Niedermayer,
arXiv:0807.0030 [hep-lat].

\bibitem{Borsanyi:2012xf} 
S.~Borsanyi, Y.~Delgado, S.~Durr, Z.~Fodor, S.~D.~Katz, S.~Krieg, T.~Lippert and D.~Nogradi {\it et al.},
arXiv:1204.4089 [hep-lat].

\bibitem{Egri:2006zm} 
G.~I.~Egri, Z.~Fodor, C.~Hoelbling, S.~D.~Katz, D.~Nogradi and K.~K.~Szabo,
Comput.\ Phys.\ Commun.\  {\bf 177}, 631 (2007)
[hep-lat/0611022].

\end{thebibliography}
\end{document}